\documentclass[twocolumn,twoside,slac_two]{revtex4}
\usepackage{fancyhdr}
\usepackage{graphicx}
\usepackage{subfigure}
\usepackage{mathptmx}
\usepackage{helvet}
\usepackage{array}
\usepackage{multirow}
\usepackage{amsmath}
\usepackage{amssymb}
\usepackage{latexsym}
\usepackage{booktabs}
\usepackage{mathrsfs}

\begin{document}

\title{Search for Contact Interactions in the Dimuon Final State at ATLAS}

%

\author{E. N. Thompson, S. Willocq}
\affiliation{University of Massachusetts, Amherst}

\author{K. M. Black}
\affiliation{Harvard University}

\begin{abstract}
The Standard Model has been successful in describing many fundamental aspects of particle physics. However, there are some remaining puzzles that are not explained within the context of its present framework.  We discuss the possibility to discover new physics in the ATLAS Detector via a four-fermion contact interaction, much in the same way Fermi first described Weak interactions.  Using a simple ratio method on dimuon events, we can set a 95\% C.L. lower limit on the effective scale $\Lambda\nobreak=$\nobreak7.5~TeV (8.7~TeV) for the constructive Left-left Isoscalar Model of quark compositeness with $\mathscr{L} =$~100~pb$^{-1}$ (200 pb$^{-1}$) of data at $\sqrt{s}~=$~10~TeV.
\end{abstract}

\maketitle

\thispagestyle{fancy}


\section{Introduction}

New physics discovery may be just around the corner as the Large Hadron Collider (LHC) prepares for first collisions this winter.  The Standard Model has thus far shown impressive predictive power and agreement with experiment; yet the cause of Electroweak Symmetry Breaking (EWSB), necessary to give mass to the W$^{\pm}$ and Z$^0$ bosons, remains experimentally unconfirmed. In the last decades, many models outside of the current Standard Model framework have been developed to address this.

One way to model a new interaction with unknown couplings is in the form of a ``4-fermion'' contact interaction. Interactions with a dimuon final state have been chosen for this analysis, as they provide a clean signature in the early stages of understanding the ATLAS detector.

In the first year of LHC running, we expect 100~-~200~pb$^{-1}$ of data at $\sqrt{s} =$~10~TeV.  In these proceedings, we discuss the feasibility of discovering or setting limits on new physics via contact interactions with early ATLAS data. Note also that the results presented here have not yet been officially approved by the ATLAS Collaboration.

\section{Theoretical Background}

Beyond the Standard Model processes, such as large extra spacial dimmensions (LED) in the ADD model \cite{LED} or quark/lepton compositeness \cite{PDG}, may be described as a 4-fermion contact interaction.  In the same spirit as the Fermi Interaction describes $\beta$-decay without directly knowing the intermediate process \cite{fermi}, one can write an effective Lagrangian describing a new interaction \cite{lane}:
\begin{eqnarray}\label{lagrangian}
\cal L &=& \frac{g^2}{\Lambda^2}~[~\eta_{LL}\left(\bar{\psi}_L\gamma^{\mu}\psi_L\right)\left(\bar{\psi}'_L\gamma_{\mu}\psi'_L\right) \nonumber \\
&+& \eta_{RR}\left(\bar{\psi}_R\gamma^{\mu}\psi_R\right) \left(\bar{\psi}'_R\gamma_{\mu}\psi'_R\right) \nonumber \\
&+& \eta_{LR}\left(\bar{\psi}_L\gamma^{\mu}\psi_L\right)\left(\bar{\psi}'_R\gamma_{\mu}\psi'_R\right) \nonumber \\
&+& \eta_{RL}\left(\bar{\psi}_R\gamma^{\mu}\psi_R\right)\left(\bar{\psi}'_L\gamma_{\mu}\psi'_L\right)~]~~,
\end{eqnarray}

\noindent where g is a coupling constant, and $\psi_{L,R}$ and $\psi_{L,R}'$ are the incoming and outgoing left and right fermionic fields, respectively.  The interaction appears experimentally as a deviation from the SM dilepton mass spectrum, which originates from Drell-Yan (DY) production ($q\bar{q}\rightarrow\gamma,Z\rightarrow l^{+}l^{-}$).  The value of $\eta$ is the sign of the interaction term; there can be either constructive ($\eta=-1$) or destructive ($\eta=+1$) interference with the DY process.

\begin{figure}[ht]
\begin{picture}(100,40)
\thicklines
\put (6,10){\line(1,1){28}}
\put (11,15){\vector(-1,-1){0}}
\put (28,32){\vector(1,1){0}}
\put (33,10){\line(-1,1){28}}
\put (13,30){\vector(1,-1){0}}
\put (26,17){\vector(-1,1){0}}
\put (6,29){$q$}
\put (6,17){$\bar{q}$}
\put (31,29){$q'$}
\put (31,17){$\bar{q}'$}
\put (16.5,0){($a$)}

\put (51,10){\line(1,1){28}}
\put (56,15){\vector(-1,-1){0}}
\put (73,32){\vector(1,1){0}}
\put (78,10){\line(-1,1){28}}
\put (58,30){\vector(1,-1){0}}
\put (71,17){\vector(-1,1){0}}
\put (51,29){$q$}
\put (51,17){$\bar{q}$}
\put (76,29){$l^-$}
\put (76,17){$l^+$}
\put (61.5,0){($b$)}

\end{picture}

\caption{Representative Feynman diagrams for various 4-fermion contact interactions. Examples of possible production at the LHC are shown for ($a$) dijet production and ($b$) dilepton production. \label{contacts}}
\end{figure}
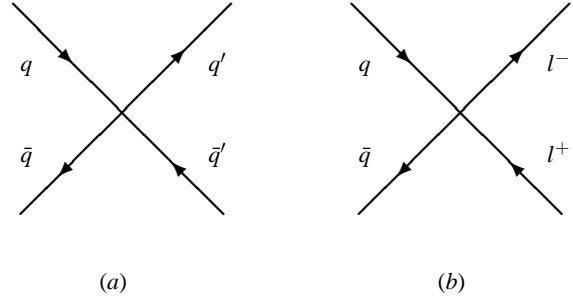

\indent Some examples of contact interaction diagrams are shown in Fig. \ref{contacts}.  In the case of Fig. \ref{contacts}b, the ``contact'' can be understood as being between incoming partons and final-state charged leptons at some scale $\Lambda$.  More specifically, the Lagrangian for the 4-fermion contact interaction with a dimuon final state ($qq\mu\mu$) is given by:
\begin{eqnarray}\label{fermilagrangian}
\cal L &=& \frac{g^2}{\Lambda^2}[~\eta_{LL}\left(\bar{q}_L\gamma^{\mu}q_L\right)\left(\bar{\mu}_L\gamma_{\mu}\mu_L\right)  \nonumber \\
&+& \eta_{RR}\left(\bar{q}_R\gamma^{\mu}q_R\right) \left(\bar{\mu}_L\gamma_{\mu}\mu_L\right)  \nonumber \\
&+& \eta_{LR}\left(\bar{q}_L\gamma^{\mu}q_L\right)\left(\bar{\mu}_R\gamma_{\mu}\mu_R\right)  \nonumber \\
&+& \eta_{RL}\left(\bar{q}_R\gamma^{\mu}q_R\right) \left(\bar{\mu}_L\gamma_{\mu}\mu_L\right)~]~~.
\end{eqnarray}

\noindent Here, $q_{L,R}$ are the left- or right-handed quark doublets, $u_{L,R}$ and $d_{L,R}$ are the left or right quark singlets, and $\mu_{L,R}$ are the left- or right-handed muon singlets.

\begin{figure}[ht]
\begin{center}
\includegraphics[width=0.9\linewidth]{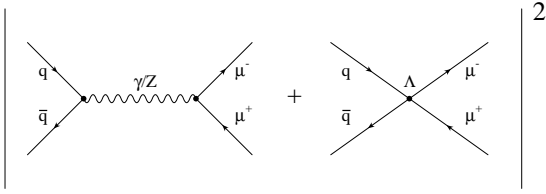}
\end{center}
\vspace{-.4cm}
\caption{Production mechanism of Drell-Yan with additional contact term with scale $\Lambda$ in the dimuon final state.}
\label{feynman}
\end{figure}

The total cross section is then given by the general form
\begin{equation}\label{crosssection}
\sigma = DY + \frac{I}{\Lambda^{2}} + \frac{C}{\Lambda^{4}}~~~,
\end{equation}
with a $DY$ term (the cross-section of the Standard Model Drell-Yan spectrum), an interference term $I$ which goes as $\Lambda^{-2}$, and the contribution of the contact interaction $C$ which goes as $\Lambda^{-4}$.

\section{Contact Interaction Analysis Method}

\begin{figure}[ht]
\begin{center}
\includegraphics[width=1\linewidth]{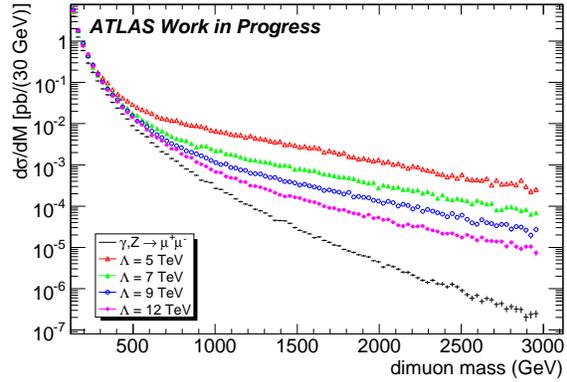}
\end{center}
\vspace{-.8cm}
\caption{Dimuon invariant mass spectra for various $\Lambda$ values. Note that the signal distribution tends to the DY shape as $\Lambda\rightarrow\infty$.}
\label{lambdas}
\end{figure}

Monte Carlo event generation using PYTHIA~\cite{pythia} has been done for the parity-conserving Left-left Isoscalar Model (L-LIM) of fermion compositeness, corresponding to the first term in Eq.~(\ref{fermilagrangian}). We chose four benchmark signal values to calculate the expected limit: $\Lambda$ = 5, 7, 9 and 12~TeV.  In the event generation stage, the following selection criteria were used:
\begin{itemize}
\item two or more final state muons from the hard scattering
\item dimuon invariant mass M$_{\mu\mu}>120$~GeV
\item transverse momentum $p_T~>~5$~GeV and pseudo-rapidity $|\eta|~<~2.8$ for final state muons.
\end{itemize}
\noindent These requirements were chosen in order to increase statistics in the signal region and to be within the geometrical acceptance of the Muon Spectrometer. In Table \ref{gen}, we show the production cross-section times the muon branching fraction ($X\rightarrow\mu\mu$) for each of the benchmark $\Lambda$ values. Detector response was simulated using the standard ATLAS fast simulation, and the resulting dimuon invariant mass was calculated.  We then multiply by a mass dependent k-factor (ranging from 1.31 to 1.15 from 120 GeV to 2000 GeV \cite{csc}) and a dimuon reconstruction efficiency factor of (0.85)$^2$ = 0.07225. After event generation and fast simulation, muon candidates are chosen to be within pseudorapidity $|\eta|~<$~2.5 (the geometrical acceptance of the Inner Detector \cite{detector}) and dimuons are required to have an invariant mass $M_{\mu\mu}~>$~120~GeV.  Fig.~\ref{lambdas} shows the resulting dimuon differential cross-section in the constructive Left-left Isoscalar Model. The DY spectrum corresponds to $\Lambda\rightarrow\infty$.\\
\vspace{-.4cm}
\begin{table}[ht]
\centering
\caption{Cross-sections for benchmark $\Lambda$ values in the constructive Left-left Isoscalar Model for pp collisions at $\sqrt{s}=10$~TeV  (for $M_{\mu\mu} >$ 120 GeV). \label{gen}}
\begin{tabular}{|c|c|}
\hline
$\Lambda$ (TeV) & $\sigma \times bf$ [pb] \\ 
\hline
5 & 13.28 \\
7 & 12.85 \\
9 & 12.75 \\
12 & 12.54 \\
$\infty$ (DY) & 12.52 \\
\hline
\end{tabular}
\end{table}

\subsection{Limit Setting and Discovery Potential Using the Ratio Method}

One method to look for an excess of events in the high-mass DY spectrum is to simply count the number of events above an invariant mass threshold ($M_0$) and divide by the number of events in a control region where the expected shape and cross-section of the DY are tightly constrained from previous experiments:

\begin{equation}\label{ratio}
R_{\Lambda,DY,d} = \frac{N_{M>M_0}}{N_{M<M_0}}~~.
\end{equation}	
\\
\noindent Note that $R_{\Lambda\rightarrow\infty} = R_{DY}$, the expected ratio from the DY invariant mass spectrum.  The number of events in any region is given by $N~=~\mathscr{L}\nobreak\epsilon\nobreak\sigma\nobreak bf(X\nobreak\rightarrow\nobreak\mu\mu)$, where $\mathscr{L}$ is the integrated luminosity,  $\epsilon$ is the acceptance and reconstruction efficiency, $\sigma$ is the differential cross-section from Eq~(\ref{crosssection}), and ($bf(X\nobreak\rightarrow\nobreak\mu\mu)$) is the branching fraction in the dimuon channel. When taking the ratio $R_{\Lambda,DY}$ from Monte Carlo simulations or $R_{d}$ from collected data, the uncertainty in the luminosity cancels in the ratio.  The effect of strongly-correlated uncertainties between the two regions is also reduced, which is especially advantageous in the first months of data taking, when the uncertainties are expected to be large. \\
\indent The ratio in Eq. (\ref{ratio}), expected for the DY-only spectrum, can be compared with an expected signal ratio using the following significance calculation:

\begin{equation}
 S_{lim} = \frac{R_{\Lambda} - R_{DY}}{\sqrt{\sigma_{\Lambda}^2 + \sigma_{DY}^2} }~~.
\end{equation} 
\\
Here $\sigma_\Lambda$ is the statistical uncertainty on the signal ratio, assuming that new physics did exist but had fluctuated downwards and was missed.  Systematic uncertainties on the background ratio are incorporated in $\sigma_{DY}$, which include detector effects and our limited knowledge of Standard Model parton density functions, backgrounds, etc.  In calculating the actual limit, we would use real data in place of the DY-only Monte Carlo sample for comparison to the ratio expected for a given $\Lambda$ scale.  The $\Lambda$ scale which results in a significance of $S_{lim}~=~1.96$ corresponds to a 95\% confidence level limit.

\subsection{Expected Reach With Early Data at $\sqrt{s} =$ 10 TeV}

\indent The effect of systematic uncertainties on the significance calculation was determined for the DY background. It was found that muon resolution, muon efficiency, and k-factor uncertainties were the largest contributors to the overall ratio uncertainty, as they are momentum-dependent.  The efficiency uncertainty was chosen to rise from 1\% to 15\% over the mass range of 120~GeV-2000~GeV, the resolution uncertainty calculation was based on a mis-alignment the Muon Spectrometer four times worse than design, and the k-factor uncertainty was based on minimizing and maximizing the slope of the mass-dependent k-factor functional form within errors previously determined \cite{csc}.  Note that the values quoted are an exaggeration of expected first-year systematics.  Table \ref{systematics} summarizes these results.

\begin{table}[ht]
\centering
\caption{Expected systematics for the first months of ATLAS running. $N_{high}$ and $N_{low}$ are the expected number of events for 100~pb$^{-1}$ above and below $M_0$, respectively.\label{systematics}}
\begin{tabular}{|l|c|c|c|c|}
\hline
Mass cut @ 450 GeV & $N_{low}$ & $N_{high}$ & DY ratio & \% diff \\
\hline
\hline
Nominal SM         & 660.5    & 7.6       & 0.0115   &        \\
\hline
Resolution         & 732.7    & 9.0       & 0.0123   & 6.7\%  \\
p$_T$ scale (1\% up) & 699.3  & 7.9       & 0.0113   & -1.5\% \\
p$_T$ scale (1\% down) & 632.5 & 7.3      & 0.0116   & 1.1\%  \\
Efficiency (up to 15\%) & 669.5 & 7.8     & 0.0117   & 2.0\%  \\
Efficiency (down 15\%)  & 651.4 & 7.3     & 0.0112   & -2.1\%  \\
k-factor (max slope)    & 635.3 & 7.4     & 0.0117   & 1.9\% \\ 
k-factor (min slope)    & 700.5 & 7.8     & 0.0111   & -3.1\% \\
\hline
\end{tabular}
\end{table}

\begin{figure}[ht]
\begin{center}
\includegraphics[width=.9\linewidth]{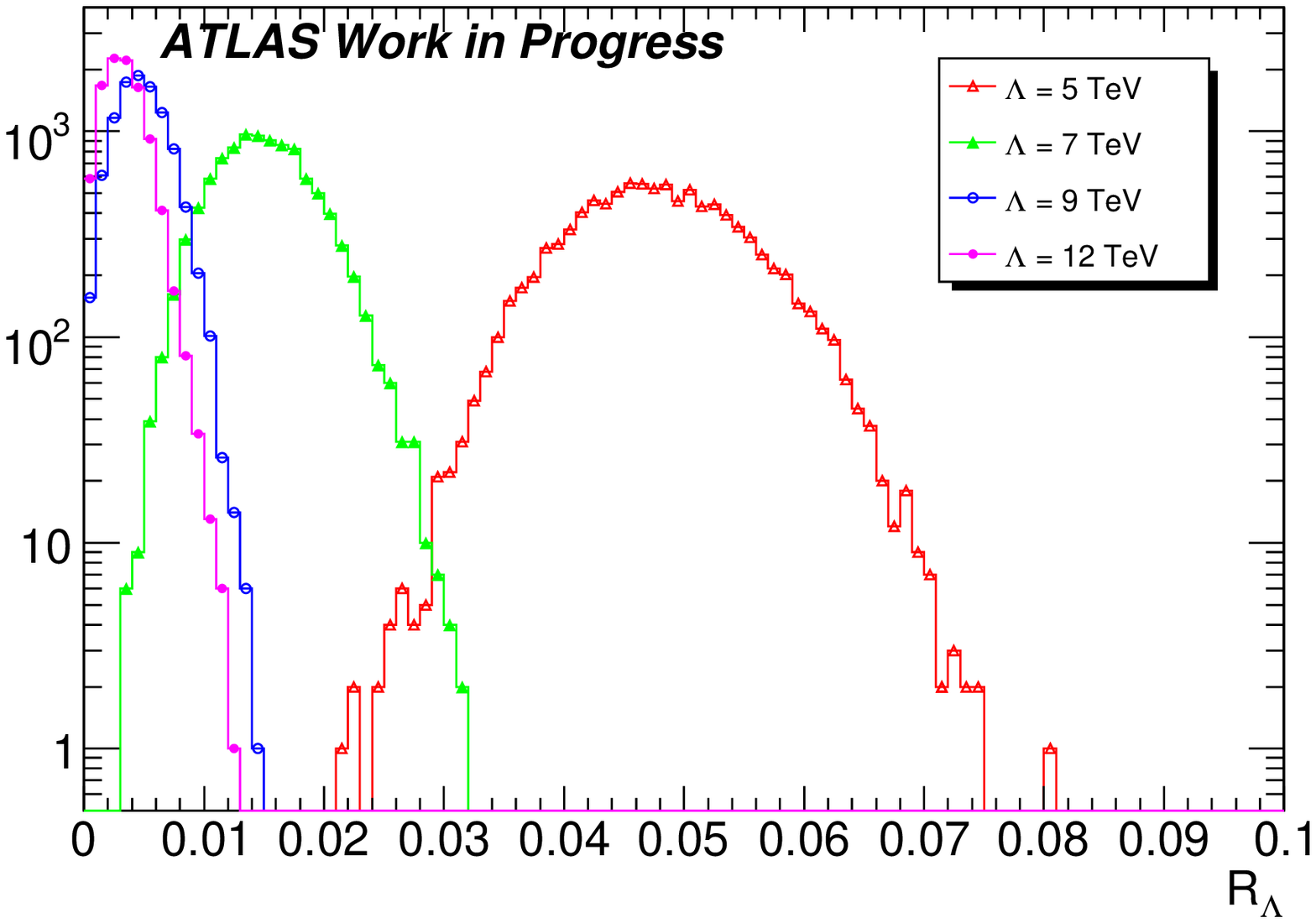}\\
\includegraphics[width=.9\linewidth]{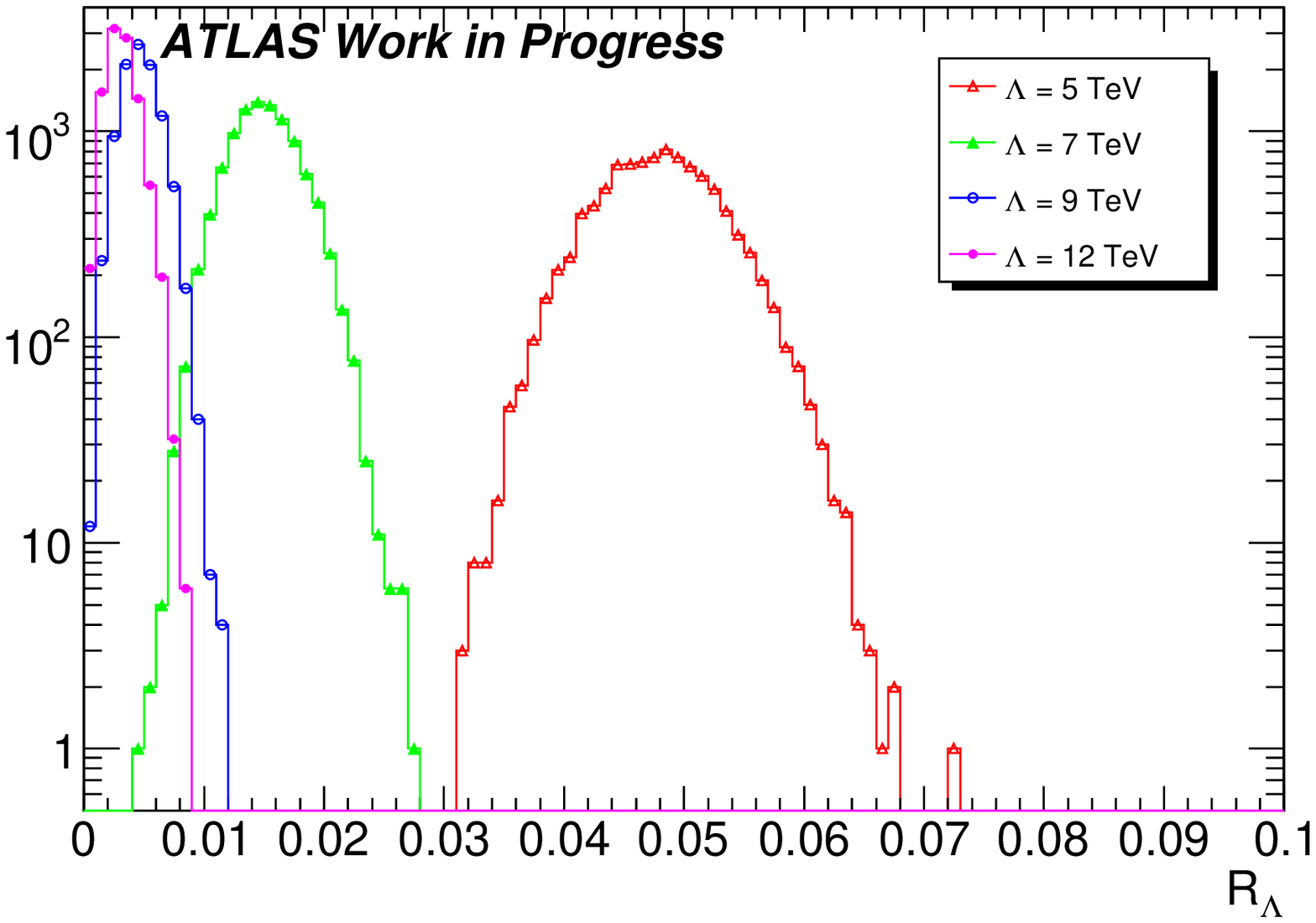}
\end{center}
\vspace{-.8cm}
\caption{Comparing various $R_{\Lambda}$ distributions for $\mathscr{L} =$ 100 pb$^{-1}$ (top) and 200 pb$^{-1}$ (bottom).}
\label{pseudo}
\end{figure}

To determine the impact of statistical uncertainty in the signal ratios ($\sigma_{\Lambda}$), the RMS of $R_{\Lambda}$ distributions from performing 10000 pseudo-experiments were calculated on each of the benchmark values (Fig.~\ref{pseudo}, Table~\ref{ratio_table}). The number of sample events ($N_i$) chosen for the $i^{th}$ pseudo-experiment was determined from a Poisson distribution about the expected number of events $N_{expected}$ in 100~pb$^{-1}$ (or 200~pb$^{-1}$) of data.  Each pseudo-experiment was then created by randomly sampling $N_i$ events from a dataset corresponding to hundreds of~fb$^{-1}$.

\begin{figure*}[ht]
\begin{center}
\includegraphics[width=.40\linewidth]{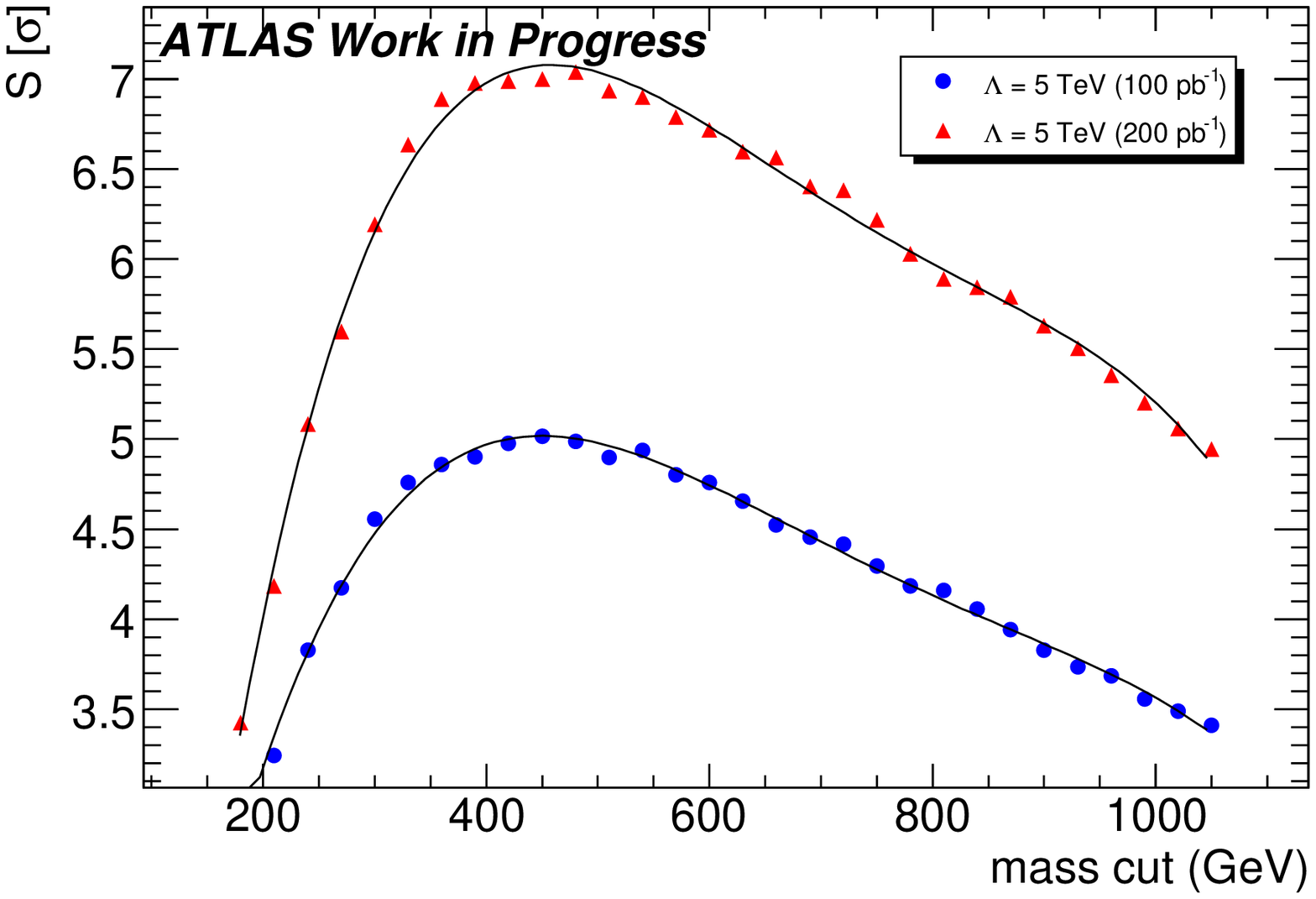}
\includegraphics[width=.40\linewidth]{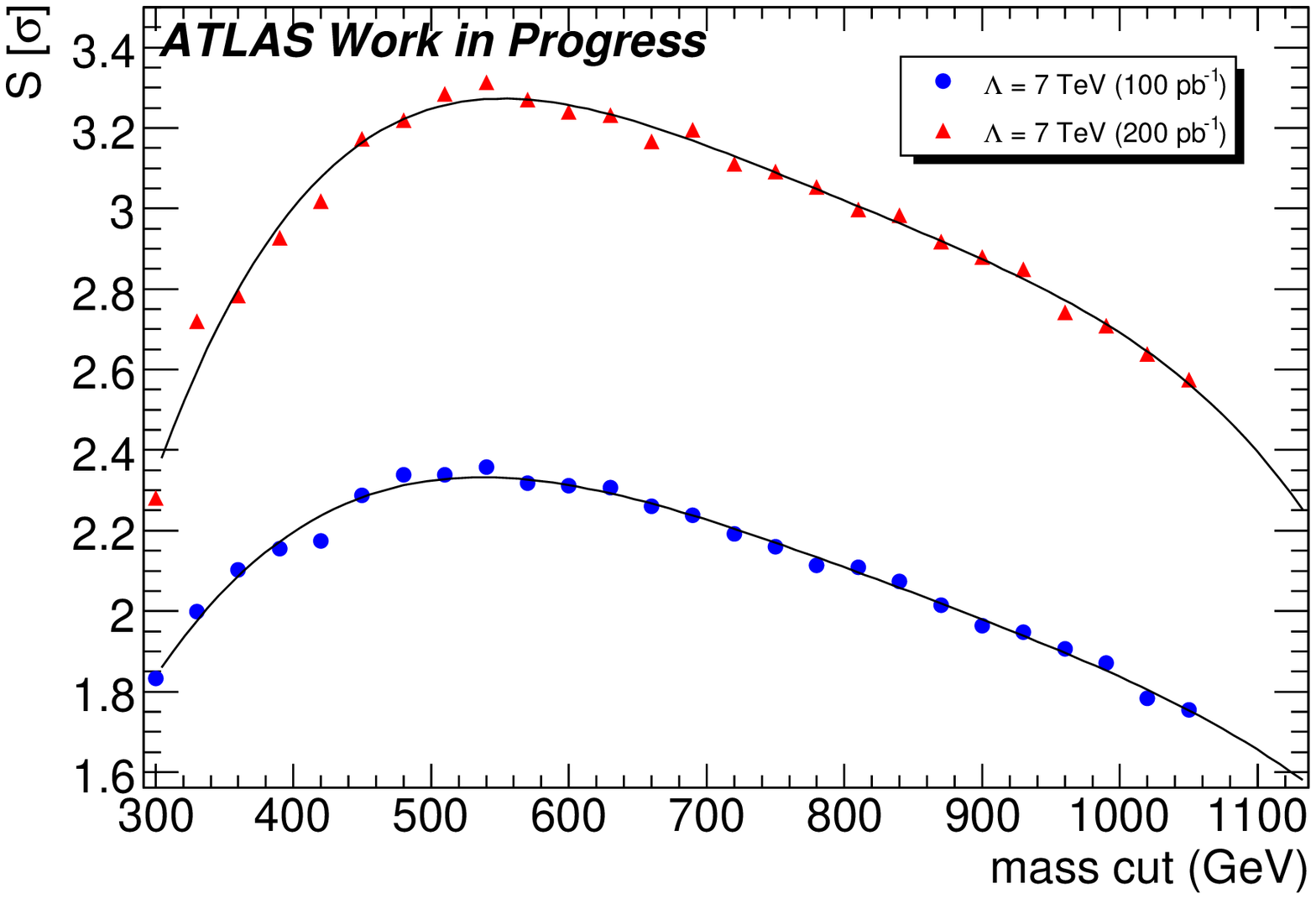} \\
\includegraphics[width=.40\linewidth]{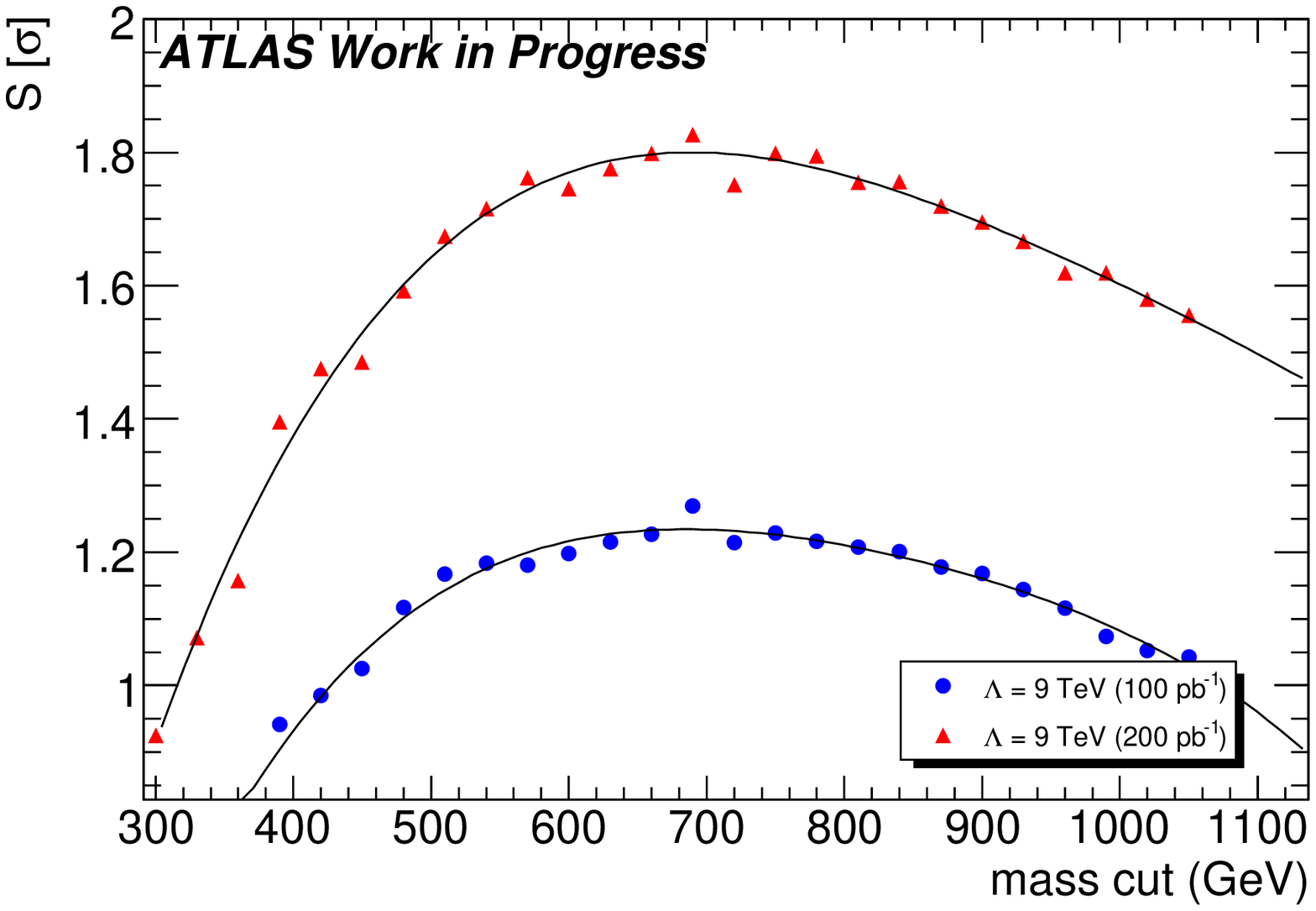} 
\includegraphics[width=.40\linewidth]{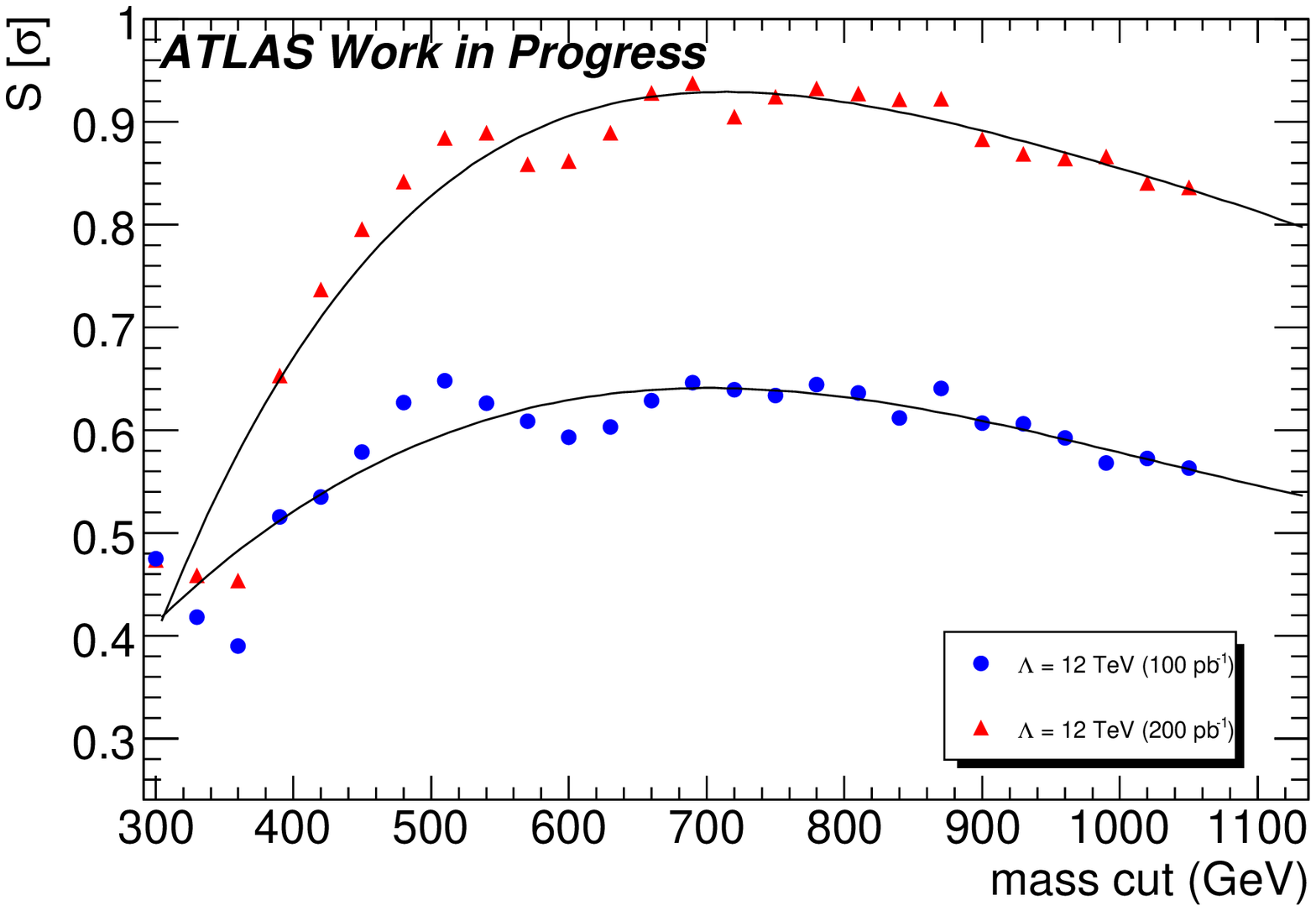}
\end{center}
\vspace{-.4cm}
\caption{Distributions of significance ($S_{lim}$) by varying $M_0$ for different $\Lambda$ values, with the blue curve corresponding to 100 pb$^{-1}$ and the red curve corresponding to 200 pb$^{-1}$.}
\label{m0max}
\end{figure*}

The value of the expected ratios $R_{DY, \Lambda}$ are highly dependent on the choice of $M_0$, the cutoff between the high and low dimuon invariant mass regimes.  To account for this, we calculate $S_{lim}$ as a function of $M_0$ (Fig.~\ref{m0max}).  The $M_0$ which maximizes $S_{lim}$ for each $\Lambda$ value is used when setting the limit on~$\Lambda$ (see Table \ref{ratio_table}).  We have found that the maximum is very broad in $M_0$, and that increasing the luminosity does not affect the chosen mass cut significantly.

\begin{table}[ht]
\centering
\caption{Ratios $R_{DY}$ and $R_{\Lambda}$ are shown for each $\Lambda$ with corresponding best $M_0$ cut in 100 pb$^{-1}$.\label{ratio_table}}
\begin{tabular}{|c|c||c|c||c|c|c|}
\hline
\multirow{2}{*}{$\Lambda$} & \multirow{2}{*}{Best $M_0$ } & \multicolumn{2}{c||}{Drell-Yan only} & \multicolumn{3}{c|}{Drell-Yan + signal} \\ [4pt]
\cline{3-7}  & & $\frac{N_{high}}{N_{low}}$ & $R_{DY}$ & $\frac{N_{high}}{N_{low}}$  & $R_{\Lambda}$ & $\sigma_{\Lambda}$ \\ [4pt]
\hline
5 TeV & 450 GeV & $\frac{10.20}{805.52}$ & 0.0127 & $\frac{44.33}{820.95}$& 0.0541 & 0.0082\\[3pt]

7 TeV & 540 GeV & $\frac{4.98}{809.37}$ & 0.0062 & $\frac{13.66}{808.37}$& 0.0170 & 0.0046 \\[3pt]

9 TeV & 750 GeV & $\frac{1.27}{812.36}$ & 0.0016 & $\frac{3.73}{820.56}$ & 0.0046 & 0.0024 \\[3pt]

12 TeV & 780 GeV & $\frac{1.07}{812.53}$ & 0.0013 & $\frac{2.00}{805.93}$ & 0.0025 & 0.0017 \\[3pt]
\hline
\end{tabular}
\end{table}

\begin{figure}[ht]
\begin{center}
\includegraphics[width=0.95\linewidth]{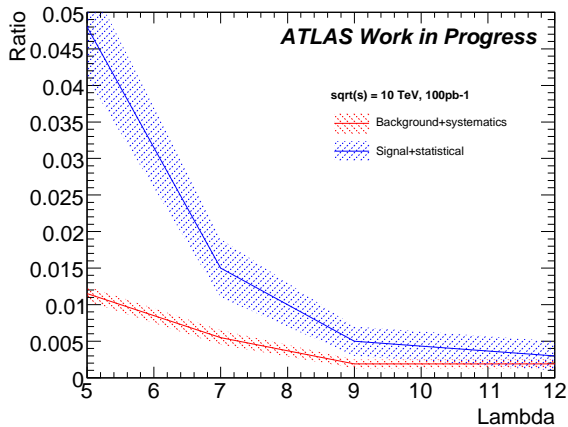}
\end{center}
\vspace{-.8cm}
\caption{Ratio versus Lambda, using $M_0$ for maximized significance.}
\label{ratios}
\end{figure}

\begin{figure}[ht]
\begin{center}
\includegraphics[width=1\linewidth]{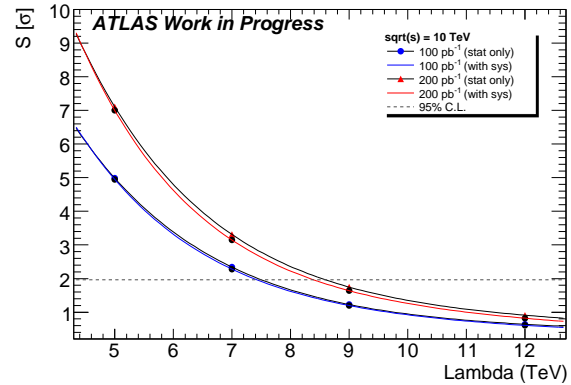}
\end{center}
\vspace{-.8cm}
\caption{The expected 95\% C.L. limit on $\Lambda$ corresponds to a significance ($S_{lim}$) of 1.96$\sigma$, shown here for 100~pb$^{-1}$ and 200~pb$^{-1}$.}
\label{limit}
\end{figure}

Note that in early data, while the resolution uncertainty is the largest systematic, statistical uncertainty dominates in calculating the ratio (Fig. \ref{ratios}). After fitting the significance for the various $\Lambda$ values (5, 7, 9 and 12 TeV), we expect to be able to set a new limit on the constructive L-LIS model of fermion compositeness at $\Lambda~=$~7.5~TeV with 100~pb$^{-1}$, or $\Lambda~=$~8.7~TeV with 200~pb$^{-1}$ (Fig.~\ref{limit}).

\newpage


\end{document}